\documentclass[prd,twocolumn,aps]{revtex4}

\newcommand{\beq}{\begin{equation}}
\newcommand{\eeq}{\end{equation}}
\newcommand{\bea}{\begin{eqnarray}}
\newcommand{\eea}{\end{eqnarray}}
\newcommand{\lv}{\langle}
\newcommand{\rv}{\rangle}
\newcommand{\ov}{\overline}

\usepackage{graphicx}

\begin{document}

\title{Supersymmetric Grand Unification: the quest for the theory
\footnote{Based on talks given in BW2003
Workshop on Mathematical, Theoretical and Phenomenological Challenges Beyond the Standard Model: Perspectives of Balkans Collaboration, Vrnjacka Banja, Serbia, 29 Aug - 2 Sep 2003.}}
\author{Alejandra Melfo$^{(1)}$, 
Goran Senjanovi\'c$^{(2)}$}

\affiliation{$^{(1)}$ {\it Centro de F\'{\i}sica  Fundamental, Universidad de
Los Andes, M\'erida, Venezuela }}
\affiliation{$^{(2)}${\it International Centre for Theoretical Physics,
34100 Trieste, Italy }}

\begin{abstract} 
With the advent of neutrino masses, it has become more and more acknowledged
that SO(10) is a more suitable theory than SU(5): it leads naturally to small
neutrino masses via the see-saw mechanism,
it has a simpler and more predictive Yukawa sector. There is
however  a rather strong disagreement on what the minimal consistent SO(10)
theory is, i.e. what the Higgs sector is. The issue is particularly sensitive
in the context of low-energy supersymmetry.

\end{abstract}


 \maketitle

\section{Introduction}
 
 Supersymmetric Grand Unification has been one of the main extensions of the
Standard Model (SM) for now more than two decades. Today, however, it is in search of a
universally accepted minimal, consistent model.  With the growing evidence for
neutrino masses \cite{neutrino}, it is becoming more and more clear that the
SU(5) theory is not good
enough: it contains too many parameters in the Yukawa sector. The situation is
much more appealing in the SO(10) scenario, which is custom fit  to explain
small neutrino masses in a simple and fairly predictive manner.  The main
dispute lies in the breaking of SO(10) down to the Minimal Supersymmetric
Standard Model (MSSM), in the delicate question of the choice of the Higgs
superfields.

Roughly speaking, there are two schools of thought: one that sticks to the
small representations, which guarantees asymptotic freedom above $M_{GUT}$, but
must make  use of  higher dimensional operators, suppressed by $M_{Pl}$;
one that argues in favor of the renormalizable theory only, even at the
price of becoming strong between $M_{GUT}$ and the Planck scale. Each program
has its pros and cons.  The first one in a sense goes beyond grand unification
by appealing to the string picture in order to provide additional horizontal
symmetries needed to simplify the theory plagued by many couplings. The second
one is based on pure grand unification, with the hope that the Planck scale
physics plays a negligible role.  It is the second one that we discuss at
length in this talk. 

\section{Why grand unification and why supersymmetry ?}

No excuse needs to be offered for the natural wish to unify the strong and
electro-weak interactions.  This appealing idea has two important generic
features: proton decay and the existence of magnetic monopoles. They are by
themselves sufficient reason to pursue the unification scenario. 

There are three important reasons to incorporate low-energy supersymmetry in
this program:
{\it i)} the hierarchy problem of the Higgs mass, {\it ii)} the gauge coupling
unification, and {\it iii)} the Higgs mechanism in the form of radiative
symmetry breaking.  Let us briefly discuss them.
\begin{itemize}
\item  Supersymmetry per se says nothing about the smallness of the
Higgs mass (the hierarchy problem), it just keeps the perturbative effect small,
the way that chiral symmetries protect small Yukawa couplings.  The old
feelings that this might not be such a big deal, since the cosmological
constant does not get protected in a similar way, are becoming more widespread
today. 
\item  Gauge coupling unification of the MSSM \cite{susyunif} is a rather
remarkable
phenomenon, but its meaning is not  completely clear. Namely, if one believes
in a desert between $M_{W}$ and $M_{GUT}$, then this becomes a crucial
ingredient. The desert is a property of the minimal gauge group, SU(5), which
is not a good theory of neutrino masses. In SO(10), on the other hand,
supersymmetry is not essential at all; the theory works even better without
supersymmetry since then it predicts intermediate scales in the range
$10^{10}-10^{14} GeV$, ideal for neutrino masses via the see-saw mechanism
\cite{seesaw},
and for leptogenesis \cite{Fukugita:1986hr}.

It is worth stressing though that  supersymmetric grand unification was
anticipated already  in 1981,  and it gave a rationale for a heavy top quark
with a mass around $200 GeV$ (needed to increase the $\rho$ parameter and help
change $\sin^{2}\theta_{W}$ from its then accepted value of $0.20$ to the
current $0.23$, see e.g. last paper in Ref.\cite{susyunif}).

\item  Radiative symmetry breaking and the Higgs mechanism \cite{Alvarez-Gaume:1983gj}. The
tachyonic property of the Higgs mass term has bothered people for a long time.
It is of course purely a question of taste, for either sign of $M_{H}^{2}$ is
equally probable. Since the charged sfermion mass terms are definitely not
tachyonic, in supersymmetry one could ask why is the Higgs scalar so special.
The answer is rather simple: if the nature if the scalar mass terms is
determined by some large scale, and if all $m^{2} > 0$, it turns out that the
Higgs doublet coupled to the top quark rather naturally becomes tachyonic at
low scales due to the large top Yukawa coupling. This was kind of prophetic
more than twenty years ago, and it could be a rationale for such a heavy top
quark. Admittedly a little fine-tuning is still needed between the so-called
$\mu$ term and the stop mass, but wether this is a small or a large problem is
still disputable.
\end{itemize}

Suppose we accept low energy supersymmetry as natural in grand unification.
We then face the task of identifying the minimal consistent supersymmetric
grand unified theory and then hope that it will be confirmed by experiment.
Due to the miraculous gauge coupling unification in the MSSM, we are then
tempted to accept the idea of the desert.  Since the desert is a natural
property of SU(5),  it is not surprising that SU(5) was considered for a long
time the main candidate for a supersymmetric GUT.  Why not stick to this idea?

The point is that SU(5), at least in its simplest form, favors massless
neutrinos.  Higher dimensional operators can only provide small neutrino
masses, not large enough to explain $\Delta m_{\odot}^{2} \simeq 10^{-5} eV
^{2}$ and especially $\Delta m_{A}^{2} \simeq 10^{-3} eV
^{2}$. While we can try to remedy this in one way or another, the fact is that
the realistic theory needs too many parameters to be predictive. Let us
try to justify this claim by analyzing $SU(5)$ in some detail.

\section{Supersymmetric SU(5)}

The minimal Higgs sector needed to break the symmetry completely down to
$U(1)_{em}\times SU(3)_{c}$ consists of the adjoint $24_{H}$ and two
fundamentals $5_{H}$ and $\bar 5_{H}$. The Higgs superpotential is quite
simple
\begin{equation}
W_{H} = m (24_{H^{2}}) + \lambda (24_{H})^{3} + \mu \bar 5_{H} \,5_{H} + \alpha
5_{H} \, 24_{H} \, \bar 5_{H}
\label{WHsu5}
\end{equation}
and so is the Yukawa one
\beq
W_{Y} = y_{d}\, 10_{F}\, \bar 5_{F} \, \bar 5_{H} + y_{u} \, 10_{F} \,10_{F}
\, 5_{H}
\label{Wysu5}
\eeq
since the charged fermions belong to $\bar 5_{F}$ and $10_{F}$.
 
 The above theory is usually called the minimal supersymmetric SU(5) theory.
It apparently has a small number of parameters:
\begin{itemize}
 \item 3 real $y_{d}$ (after diagonalization)
\item 2 x 6 = 12 real $y_{u}$ ($y_{u}$ is a symmetric matrix in generation
space)
\item 2 real $\mu$, $m$ (after rotations)
\item 2 x 2 = 4 real $\lambda$, $\alpha$
 \end{itemize}
 In total, 21 real parameters. The trouble is that the theory fails badly \cite{Chanowitz:1977ye}.
Neutrinos are massless, and thus $V_{\ell} = 1$; furthermore, the
relations $m_{e} = m_{d}$ at $M_{GUT}$ fail, except for the third generation.
The most conservative approach would be to blame the failure  in the absence
of higher dimensional operators, this way no change in the structure of the
theory is needed.

Neutrino masses are then given by the Weinberg-type operator
\beq
W_{y_{\nu}} = y^{\nu} \frac{\bar 5_{F} \bar 5_{F} 5_{H} 5_{H}}{M_{Pl}}
\label{Wnusu5}
\eeq
 giving
 \begin{itemize} \item  6 x 2 = 12 new real parameters in $y_{\nu}$.
\end{itemize}
 
 Similarly, we must add higher dimensional operators to (\ref{Wysu5}),
 \bea
 \Delta W_{y} &=& \frac{ (24_{H}\, 5_{H})}{M_{Pl}} \,10_{F}\, 10_{F}  
 + \frac{ (10_{F}\, 10_{F}\,24_{H})}{M_{Pl}} \, 5_{H} \nonumber \\
 & & + \frac{ (24_{H}\, 5_{H})}{M_{Pl}} 10_{F \bar 5_{F}} +
 \frac{ (24_{H}\, \bar 5_{H})}{M_{Pl}} 10_{F} \, 5_{H}
 \label{DWysu5}
 \eea
 (we have omitted SU(5) indices, and represented contraction with indices in
$24_{H}$ with a bracket). This means 
\begin{itemize} 
\item 9 x 2 x 2 + 6 x 2 x 2 = 60 real parameters
\end{itemize}
and then most predictability is gone. 

What remains? First, $m_{b} = m_{\tau}$ is still there, still a success. Gauge
couplings unify as we know, but the GUT scale is not predicted precisely as
often claimed in the literature. The point is that, for the sake of
consistency, one should add higher dimensional operators to (\ref{WHsu5}), so
that one expects
\beq
\Delta W_{H} = c_{1} \frac{Tr 24_{H}^{4}}{M_{Pl}} + c_{2} \frac{(Tr
24_{H}^{2})^{2}}{M_{Pl}}
\label{DWHsu5}
\eeq
and if the coupling $\lambda$ in (\ref{WHsu5}) were to be small, these terms
would become important. But $\lambda$ is a Yukawa type coupling, i.e. it is
self-renormalizable, so it can be naturally small. This point is worth
discussing further.

At the renormalizable tree level, one gets the same masses for the color octet
and the $SU(2)_{L}$ triplet in $24_{H}$: $m_{8}= m_{3}$. This is almost always
assumed when the running from $M_{GUT}$ to $M_{W}$ is studied. Now, if
$\lambda$ is small, the $c_{i}$ terms in (\ref{DWHsu5}) can dominate; if so,
one gets $m_{3} = 4 m_{8}$. This fact alone suffices to increase $M_{GUT}$ by
an order of magnitude above the usually quoted value $M_{GUT}^{0} \simeq
10^{16} GeV$ (calculated with $c_{i}=0$). Similarly, the masses of the colored
triplets $T$ and $\bar T$ in $5_{H}$ and $\bar 5 _{H}$ would get increased by
a factor of about 30, and the $d=5$-induced proton life-time by about
$10^{3}$.  More precisely, one obtains \cite{Bajc:2002bv}
\beq
m_{T} = m_{T}^{0} \left( \frac{m_{3}}{m_{8}}\right)^{5/2},
\label{mT}
\eeq
\beq
M_{GUT} = M_{GUT}^{0} \left( \frac{M_{GUT}^{0}}{2 m_{8}}\right)^{1/2},
\eeq
 where the superscript $^{0}$ denotes the tree-level value $m_{3}= m_{8}$. In
this case
\beq
m_{8} \simeq \frac{M_{GUT}^{2}}{M_{Pl}}
\label{m8}
\eeq
so that 
\beq
M_{GUT}^{4} \simeq (M_{GUT}^{0})^{3} M_{Pl}
\label{Msu5}
\eeq
With $M_{GUT}^{0} \simeq 10^{16} GeV$, this means
\beq
M_{GUT} \simeq 10 M_{GUT}^{0} \, , \quad m_{T}\simeq 32 m_{T}^{0}
\eeq
It should be stressed that $\lambda$ small is natural technically, as much as
a small electron Yukawa coupling. Taking $\lambda \sim  O(1)$ and ruling out
the theory would be equivalent to finding that the SM does not work with all
the Yukawa couplings being of order one.  

Taking into account non-renormalizable interactions can thus save the theory.
It is important to recall that without them, the minimal SU(5) does not make
sense anyway, predicting as it does $m_{\nu} = 0$ and $m_{d} = m_{e}$; once
this is corrected the theory is still valid.

Of course, if one prefers the renormalizable theory, one needs new states such
as $45_{H}$ (in order to correct $m_{d}=m_{e}$), or $15_{H}$ to give neutrino
masses, or three (at least two) singlet right-handed neutrinos. This
introduces even more uncertainties in the computations of $M_{GUT}$ and
$\tau_{P}$. In short, the minimal realistic supersymmetric SU(5) theory is not
yet ruled out. It is indispensable to improve the experimental limit on
$\tau_{P}$ by two-three orders of magnitude. Grand unification needs
desperately a new generation of proton decay experiments.

In the supersymmetric version of SU(5), there is yet another drawback. As much
as the MSSM, it allows for the $d=4$ proton decay through terms like
\beq
\Delta W = m\lambda ' \, 10_{F} \, \bar 5_{F} \bar 5_{F} 
\label{d=4su5}
\eeq 
which contains both 
\beq
\lambda ' \left( \nu^{c }D^{c} D^{c} + Q L D^{c}\right)
\eeq
This is a disaster (unless $\lambda ' \lesssim  10^{-12}$). A way out is
assumed through the imposition of R-parity, or equivalently matter parity
\cite{rparity}
$M
: F \to -F ,  H \to H$, where $F$ stands for the fermionic (matter) superfields
and $H$ for the Higgs ones. Grand unification ought to do better than this,
and SO(10) does it as we shall see.

In any case, SU(5) does a poor job in the neutrino sector and in the charged
fermion sector it is either incomplete \cite{Senjanovic:2005sf} or it has too many parameters.  One
would have to include extra horizontal symmetries, and this route is in some
sense beyond grand unification and often needs strings attached. If we stick
to the pure grand unification, we better move on to SO(10). 

\section{Towards unification: Pati-Salam symmetry}

Quark-Lepton unification  can be considered a first step towards the complete
SO(10) unification of a family of fermions in a single representation.  Many
interesting features of SO(10) GUTs, such as a renormalizable see-saw and
R-parity conservation, are already present in partial unification based on the
Pati-Salam \cite{patisalam} group $G_{PS} = SU(4)_{c} \times SU(2)_{L} \times
SU(2)_{R}$, so it is
 instructive to review the situation there. Later, when we turn to SO(10),
decomposition of representation under the Pati-Salam subgroup will prove to be
the most useful.

 To simplify the discussion, imagine a two-step breaking of the PS symmetry
down to the MSSM
\begin{eqnarray}
SU(2)_L\times SU(2)_R\times SU(4)_c  \nonumber \\ 
\longrightarrow SU(2)_L\times SU(2)_R\times U(1)_{B-L} \times SU(3)_c \nonumber\\
 \longrightarrow SU(2)_L\times U(1)_Y\times SU(3)_c 
\label{breaking}
\end{eqnarray}
The first steps breaks $G_{PS}$ down to its maximal subgroup, the LR
(Left-Right) group \cite{leftright}, and it is simply achieved through the vev of
and adjoint representation (the numbers in parenthesis indicate the $G_{PS}$
representations)
\beq
A  = (1,1,15)
\eeq
In turn, the breaking of the $LR$ group can be  achieved by having $SU(2)_{R}$
triplets fields,  with $B-L =2 $, acquiring a vev. Triplets will couple to
fermions and 
give a mass to right-handed neutrino, providing the see-saw mechanism at the
renormalizable level.  Right-handed doublets could also do the job, but then
non-renormalizable operators have to be invoked, which means effective
operators resulting from a new theory at a higher scale, but this theory we will
discuss explicitly in the next section. 

There is a more profound reason for preferring the triplets. They  have an
even $B-L$ number, and thus preserve matter parity as we defined above. This
in turn means R-parity is not broken at a high scale.   But then it can be
easily shown that it cannot be broken afterwards,  at the 
low energy supersymmetry breaking or 
electroweak scale. More precisely,  a spontaneous breakdown 
of R-parity through the sneutrino VEV (the only candidate) would result 
in the existence of a pseudo-Majoron with its mass inversely proportional 
to the right-handed neutrino mass. This is ruled out by the Z decay width 
\cite{Aulakh:1997ba,Aulakh:1999cd}. This fact is completely analogous to the 
impossibility  of breaking R-parity spontaneously in the MSSM, where the Majoron
is  strictly massless. 

In terms of PS representations, the LR triplets are contained in the fields
\beq
\Sigma  (3,1,10) , 
\;\bar\Sigma  (3,1,\bar{10}) , \;\Sigma_c (1,3,\bar{10}) , 
\;\bar\Sigma_c (1,3,10) 
\label{asigma}
\eeq

The matter supermultiplets are
\begin{equation}
\psi (2,1,4) , \; \psi_c (1,2,\bar 4)
\end{equation}
and the minimal light Higgs multiplet is
\begin{equation}
\phi (2,2,1)
\end{equation}

The most general superpotential for the fields (\ref{asigma}) is
\begin{equation}
W = m Tr A^2 + M Tr( \Sigma\bar\Sigma + \Sigma_c\bar\Sigma_c) +  Tr(
\Sigma A \bar\Sigma - \Sigma_c A \bar\Sigma_c )
\end{equation}
where we assume the following transformation properties under Parity
\begin{equation}
\Sigma \to \Sigma_c , \quad \bar\Sigma \to \bar \Sigma_c , \quad A \to -A
\end{equation}
We choose $A$ to be a parity-odd field in order to avoid flat
directions connecting left- and right-breaking minima.

It is straightforward to show that the SM singlets in $A$, $\Sigma_{c}$ and
$\bar\Sigma_{c}$ take vevs in the required directions to achieve the (in
principle two-step)
symmetry breaking 
\begin{equation}
<A> = M_c  \quad  <\Sigma_c>= M_R \quad <\bar\Sigma_c>= M_R
\label{avev}
\end{equation}
with
\begin{equation}
M_c \simeq M , \quad M_R \simeq \sqrt{M m}
\end{equation}

As discussed in detail in \cite{Kuchimanchi:1993jg,Melfo:2003xi}, the $SU(2)_{R}$-breaking
vev lies
in a flat direction that connect them with charge-breaking vacua.  It can be
eliminated if
the soft breaking  terms break also $SU(2)_R$. If not, one
would have to appeal to operators coming from a more
complete theory as studied in the next section.   The interesting point here
is that the breaking in the minimal model leaves a number of fields
potentially light \cite{Aulakh:1999pz}.  There is a larger, accidental $SU(3)$
 symmetry broken down to $SU(2)$ by the right-handed triplet fields, hence five
Nambu-Goldstone
bosons. But the gauge symmetry $SU(2)_R\times U(1)_{B-L}$ is broken down to
$U(1)_Y$, so that three of them are eaten, leaving us with 
states $\delta_c^{++},\bar\delta_c^{++}$ that acquire a
mass only at the scale of supersymmetry breaking. 
These states are common
in supersymmetric theories that include the Left-Right group, and have been
subject of experimental search \cite{search}.  In a similar way, a color octet in
$A$ has a mass of order $M_{R}^{2}/M_{c}$, and could in principle be light.

The unification constraints give the interesting possibility
$$10^3 GeV \leq M_R \leq 10^7 GeV \quad 10^{12} GeV
\leq M_c \leq 10^{14} GeV $$
opening up the possibility of  the LHC discovering them at the
$TeV$ scale.
  For larger
$M_R$, which would be necessary if one wants to fit neutrino masses without
additional fine-tuning,
these particles become   less accessible to experiment.
However, the large number of fields in this theory implies the loss of
perturbativity at a scale around $10 M_c$, and
non-renormalizable effects suppressed by this new fundamental
scale  can be shown to guarantee that they have comparable masses
\cite{Melfo:2003xi}.
Namely, if these effects are included, the only consistent possibility is the
single-step breaking
\begin{equation}
M_R\simeq M_c\simeq 10^{10}
GeV
\end{equation}

Surely the most interesting feature of a low scale of PS symmetry breaking is
the possibility of having 
  $U(1)_{B-L}$ monopoles,
 with mass $m_M \simeq 10 M_c$. If produced in a phase transition via
 the Kibble mechanism, the requirement that
their density be less than the critical density then implies
$M_c \leq 10^{12} GeV$.
 We see that the 
single-step breaking at $M_c\sim M_R \sim 10^{10} GeV$ (in a theory including
non-renormalizable terms)
 offers the interesting possibility of potentially detectable intermediate
 mass monopoles, as long as one manages to get rid of   the
false vacuum problem of supersymmetric theories. 

One final note about PS symmetry and neutrino masses. In LR theories the see-saw
mechanism is in general non-canonical, or type II. That is,   there is a
direct left-handed neutrino mass from the induced vev of the left-handed
triplet fields in $\Sigma$ (which we shall call $\Delta$)
\cite{Mohapatra:1980yp,Lazarides:1980nt,Wetterich:1981bx} $$
\propto <\Delta> \simeq M_W^2/M_R$$
Namely,  in non-supersymmetric theories the
symmetry allows for a coupling in the potential 
\begin{equation}
\Delta V = \lambda \Delta \phi^2 \Delta^c + M^2 \Delta^2
\label{deltaphidelta}
\end{equation}
resulting in
\begin{equation}
<\Delta > = \lambda \frac{<\phi>^2 M_R}{M^2} \simeq \lambda \frac{M_W^2}{M_R}
\end{equation}
 In supersymmetry such terms are of course not present,  but one could have
interactions with, for example, a  heavy field $S$ transforming as $(1,1,3)$
under $G_{PS}$  
\begin{equation}
W= \phi^2 S + \Delta \Delta^c S + M S^2
\label{ws}
\end{equation}
  Integrating out $S$ would then
give  a contribution
\begin{equation}
\Delta W = \frac{1}{M} \Delta \phi^2 \Delta^c
\label{nonren}
\end{equation}
producing the required small vev. 
Or one could have a couple of heavy fields  $X= (2,2,\bar 10)$ and $\bar X =
(2,2,10)$, which through terms like
\begin{equation}
W = \phi\Delta X + \phi \Delta^c \bar X + M X \bar X
\label{wx}
\end{equation}
 would give the same effect. These representations in fact are the ones
appearing in the minimal SO(10) theory of the next section.

The absence of the $S, X, \bar X$ fields  in the minimal PS theory guarantees a
type I see-saw at the supersymmetric level. Breaking of supersymmetry can
generate
 a nonvanishing but
negligible vev for $\Delta$ \cite{Aulakh:1999cd}:
\begin{equation}
<\Delta > \simeq \left(\frac{m_{3/2}}{M_c} \right)^2 \frac{m_D^2}{M_R}
\end{equation}
which contributes  by a tiny factor $(m_{3/2}/M_c)^2\leq
10^{-14}$ to the usual see-saw mass term $m_\nu \simeq
m_D^2/m_{\nu_R}$.
 In short, the minimal PS model has a clean, type I see-saw.
  
  In spite of providing only a partial unification, PS theory has interesting
features, namely potentially light states and the possibility of intermediate
monopoles, that could be a way of differentiating it form other theories
at a high scale. We are however interested in grand unification,  so let us
move on.

\section{SO(10) grand unified theory}

If not for anything else, but for the fact that matter parity is a finite gauge
rotation, SO(10) would be a better candidate for a supersymmetric GUT. But, as
is well known, it also unifies a family of fermions, has charge conjugation as
a gauge transformation, has right-handed neutrinos an through the see-saw
mechanism leads naturally to small neutrino masses. And, most important, at the
renormalizable level, if one is willing to accept large representations, it
has fewer parameters than SU(5).  We will elaborate on this point as we go
along.

The issue here, and the main source of dispute among the experts in the field,
is the choice of the Higgs sector. Before deciding on this, a comment on $d=4$
proton decay in the MSSM is in order. The basic problem is the impossibility
of distinguishing the leptonic and the Higgs doublets, both being superfields.
This persists in SU(5) where you have both $\bar 5_{F}$ and $\bar 5_{H}$
superfields. In SO(10) fermions belong to $16_{F}$ and the "light" Higgs to 
$10_{H}$. This difference should be taken seriously, and all efforts should be
made to maintain it.

Not all researchers agree on this. Certainly not the people who pursue
minimality by choosing small representations, like the set ( $45_{H}$, $16_{H}$,
$\bar 16_{H}$), in order to break SO(10) down to 
$SU(3)_{c}\times SU(2)_{L}
\times U(1)_{Y}$. This way, through $\langle 16_{H} \rangle 
= \langle 16_{H} \rangle  \neq 0 $, matter parity will be broken at $M_{GUT}$;
hence the catastrophic $d=4$ proton decay. One is then forced to postulate
extra discrete symmetries in order to save the theory. 
In any case, more flavor symmetries are needed, since both the symmetry
breaking and  the fermion masses need higher-dimensional, Planck suppressed
operators whose number is rather large (at least thirteen complex couplings in
$W_{H}$, the Higgs superpotential). The Yukawa superpotential takes the form
\bea
W_{Y}  &=& y_{10} \, 16_{F}  \, \Gamma \, 16_{F} \, 10_{H} \nonumber \\
& &
+ \frac{1}{M_{Pl}} \left[ c_{1} (16_{F} \Gamma  16_{F} ) \,
 (16_{H} \Gamma 16_{H} )    \right . \nonumber \\
 & & \left .
+ c_{2} (16_{F} \Gamma  16_{F} ) \,
 (\overline{16}_{H} \Gamma \overline{16}_{H} )  \right . \nonumber \\
 & & \left . +
 c_{3} 16_{F} \Gamma^{3} 16_{F}\, 45_{H} \, 10_{H}   \right . \nonumber \\
 & & \left .
+
 c_{4} 16_{F} \Gamma^{5} 16_{F}\, \overline{16}_{H}
  \Gamma^{3} \overline{16}_{H} \right]
  \label{Wyso1016}
\eea
At $M_{GUT}$, one arrives at the prediction
\beq
m_{b} = m_{\tau} \left( 1 + c \frac{M_{GUT}}{M_{Pl}}\right)
\eeq
which works very well as we know. 

Also, the see-saw takes the so-called type I form. From (\ref{Wyso1016}),
\beq
m_{\nu_{R}} \simeq c_{4} \frac{M_{GUT}^{2}}{M_{Pl}} \simeq 10^{12} - 10^{14} \,
GeV
\eeq
which fits nicely the light neutrino masses.
The type II contribution, obtained when $16_{H}$ gets a small vev $\sim
M_{W}$, 
\beq
M_{\nu}^{II} \simeq \frac{M_{W}^{2}}{M_{Pl}} \simeq 10^{-5} - 10^{-6} \, eV
\eeq
is too small to explain either atmospheric or  solar $\nu$ data  ( maybe
relevant for small mass splits in the case of degenerate neutrinos ). 

Once flavor symmetries are added, one can do the texture exercise and look
for the most appealing model. But this program goes beyond the scope of grand
unification. We ought to try to construct the minimal realistic supersymmetric
GUT without invoking any new physics.

\subsection{The pure renormalizable supersymmetric SO(10) }

Such a theory is easily built \cite{mc,Clark:ai,Chang:1984uy,Aulakh:2003kg} with large representations in the Higgs sector
$$ 210_{H} , \,  126_{H}  + \overline{126_{H}} , \, \overline{ 10_{H}} $$
with this content the theory is not asymptotically free any more above
$M_{GUT}$ \cite{Aulakh:2002ph},  and the SO(10) gauge couplings becomes strong at the scale
$\Lambda_{F} \lesssim  10 M_{GUT}$. The Higgs superpotential
is surprisingly simple
\bea
W_{H} & = & m_{210} (210_{H})^{2}  +  m_{126}
\overline{126_{H}} 126_{H} + m_{10} (10_{H})^{2}      \nonumber \\
 & &  
+ \lambda (210_{H})^{3}  
 + \eta 126_{H }\overline{126_{H}} 210_{H}  \nonumber \\
 & &  + \alpha 10_{H} 126_{H} 210_{H} 
 + \overline\alpha 10_{H}
\overline{126_{H}} 210_{H}
\label{WHso10}
\eea
With $\lv 210_{H} \rv \neq 0 $ and $\lv 126_{H}\rv = \lv \overline{126}_{H}
\rv \neq 0$, SO(10) gets broken down to the MSSM, and then $\lv 10_{H}\rv $
completes the job in the usual manner. 

The Yukawa sector is even more simple
\beq
W_{Y} = 16_{F} (y_{10} 10_{H} + y_{126} \overline{126}_{H}) 16_{F}
\label{Wyso10}
\eeq
with only 3 real (say) $y_{10}$ couplings after diagonalization, and 6 x 2 =
12 symmetric $y_{126}$ couplings, 15 in total. 

From the $\alpha $ and $\overline \alpha $ terms one gets
\begin{eqnarray}
W_{H} &=& ... + \alpha (2,2,1)_{10} \, (2,2,15)_{126} \,  (1,1,15)_{210} +\nonumber \\
 & &  
\overline{\alpha}
(2,2,1)_{10} \, (2,2,15)_{\overline{126}} \,  (1,1,15)_{210} + ...
\end{eqnarray}
Now, the success of gauge coupling unification in the MSSM favors a single
step breaking of SO(10), so that $\lv (1,1,15)_{210} \rv \simeq M_{GUT}$. In
other words, the light Higgs is a mixture \cite{Babu:1992ia} of (at least) $(2,2,1)_{10}$ and
$(2,2,15)_{\overline{126}}$; equivalently
\beq
\lv (2,2,15)_{\overline{126}} \rv \simeq \lv (2,2,1)_{10} \rv
\label{lighthiggs}
\eeq
Since $(2,2,15)_{\overline{126}}$ is an adjoint of $SU(4)_{c}$, being
traceless it give $m_{\ell} = - 3 m_{d}$, unlike $\lv (2,2,1)_{10}\rv$, which
implies $m_{\ell} = m_{d}$. In other words, the $\lv 10_{H}\rv $ must be
responsible for the  $m_{b }\simeq m_{\tau}$ relation at $M_{GUT}$, and the 
$\lv \overline{126}_{H} \rv$  for the $m_{\mu} \simeq  3  m_{s}$ relation at
$M_{GUT}$. In this theory, the Georgi-Jarlskog program becomes automatic.

Of course, we don't know anymore why  $m_{b }\simeq m_{\tau}$, or why $10_{H}$
dominates; admittedly a loss. But not all is lost. Since  $\lv 10_{H}\rv  =
\lv (2,2,1)\rv
$ is a Pati-Salam singlet, the difference between down quark and charged
lepton mass matrices must come purely from $\lv \overline{126}_{H} \rv$ 
\beq
M_{d} - M_{e} \propto y_{126}
\label{Md-Me}
\eeq
Suppose the see-saw mechanism is dominated by the so-called type II: this is
equivalent to neutrino masses being due to the triplet
$(3,1,\overline{10})_{126}$, with
\beq
\lv (3,1,\overline{10})_{126} \rv_{\overline{126}} \simeq
\frac{M_{W}^{2}}{M_{GUT}}
\eeq
In other words
\beq
M_{\nu} \equiv M_{\nu}^{II} \simeq y_{126}\lv (3,1,\overline{10})_{126}\rv
\eeq
or
\beq
M_{\nu} \propto M_{d} - M_{e}
\eeq

In order to illustrate the point, consider only the 
 2nd and 3rd generations. 
In the basis of diagonal $M_{e}$ we have
\beq
M_{\nu} \propto \left( \begin{array}{cc} m_{\mu} - m_{s} & \epsilon_{de} \\
\epsilon_{de} & m_{\tau} -m_{b}
\end{array}
\right)
\eeq
With a small mixing $\epsilon_{de}$, we see that a large atmospheric
  mixing  is only possible $m_{b} \simeq
m_{\tau}$\cite{Bajc:2002iw}.
In other words, the experimental fact of $m_{b} \simeq
m_{\tau}$ at $M_{GUT}$, and large $\theta_{\rm atm}$ seem to favor the type
type II see-saw.

On the other hand, 
 it can be shown, in the same approximation of 2nd and 3rd 
generations only, that  if type I   dominates  it gives a small $\theta_{\rm
atm}$
\cite{Bajc:2004fj}. This is a very interesting point, for it may make it
possible to determine the nature of the see-saw in this theory, in fact, it
can
 be shown  \cite{Bajc:2004fj} that the two types are really inequivalent.

The complete picture requires detailed studies of the  three-generation
case, and numerical studies have been performed.
  A type II see-saw is still supported \cite{Goh:2003sy}, 
with the
interesting prediction of a large $\theta_{13}$ and a hierarchical  neutrino 
mass spectrum.   A small
contribution of $120_{H}$ 
\cite{Bertolini:2004eq} or higher dimensional operators\cite{Dutta:2004wv} 
allow for better fits.  Adding CP phases still may give room for type I
 \cite{japon} (for earlier work on type I see\cite{Lavoura:1993vz}).

\subsection{Unification constraints}
 
  It is certainly appealing to have an intermediate see-saw mass scale $M_{R}$,
between $10^{12} - 10^{15} \mbox{GeV}$ or so. In the non-renormalizable case, with
$16_{H}$  and $\ov{16}_{H}$, this is precisely what happens: $M_{R}\simeq c
M_{GUT}^{2}/M_{Pl}\simeq c (10^{13} -10^{14}) \mbox{GeV}$. In the renormalizable
case, with $126_{H}$ and $\ov{126}_{H}$, one needs to perform a
renormalization group  study using unification constraints. While this is in
principle possible, in practice it is hard due to the large number of
fields. The stage has recently been set, for all the particle masses were
computed\cite{Bajc:2004xe,Fukuyama:2004xs}, and the preliminary
studies show that the situation may be under control\cite{Aulakh:2004hm}. It
is interesting that the existence of intermediate mass scales lowers the GUT
scale\cite{Bajc:2004xe,Goh:2004fy} (as was found before in models
with $54_{H}$
and $45_{H}$\cite{Aulakh:2000sn}), allowing for a possibly observable $d=6$ proton
decay. 

Notice that a complete study is basically impossible. In order to perform the
running, you need to know particle masses precisely. Now, suppose you stick to
the principle of minimal fine-tuning. As an example, you fine-tune the mass of
the $W$ and $Z$ in the SM, then you know that the Higgs mass and the fermion
masses are at the same scale
\beq
m_{H}= \frac{\sqrt{\lambda}}{g}m_{W} \, , \quad m_{f} = \frac{y_{f}}{g} m_{W}
\eeq
where $\lambda $ is a $\phi^{4}$ coupling, and $y_{f}$ an appropriate fermionic
Yukawa coupling. Of course, you know the fermion masses in the SM model, and
you know $m_{H} \simeq   m_{W}$. 

In an analogous manner, at some large scale $m_{G}$ a group $G $ is broken and
there are usually a number of states that lie at $m_{G}$, with masses
\beq
m_{i} = \alpha_{i} m_{G}
\eeq
where $\alpha_{i}$ is an approximate dimensionless coupling. Most
renormalization group studies typically argue that $\alpha_{i} \simeq O(1)$ is
natural, and rely on that heavily. In the SM, you could then take
$m_{H}\simeq m_{W}$, $m_{f}\simeq m_{W}$; while reasonable for the Higgs, it
is nonsense for the fermions (except for the top quark). 
In supersymmetry {\em all} the couplings are of Yukawa type, i.e.~
self-renormalizable, and thus taking $\alpha_{i}\simeq O(1)$ may be as wrong 
as taking all $y_{f}\simeq O(1)$. While a possibly reasonable approach when
trying to get a qualitative idea of a theory, it is clearly unacceptable when
a high-precision study of $M_{GUT}$ is called for.

\subsection{Proton decay}
As you know, $d=6$ proton decay gives $\tau_{p}(d=6) \propto M_{GUT}^{4}$,
while $(d=5)$ gives  $\tau_{p}(d=5) \propto M_{GUT}^{2}$. In view of the
discussion above, the high-precision determination of $\tau_{p}$ appears
almost impossible in SO(10) (and even in  SU(5)).  Preliminary studies
\cite{borutref} indicate fast $d=5$  decay 
as expected. 

 We are ignoring
the higher dimensional operators of order $M_{GUT}/M_{Pl} \simeq 10^{-2} -
10^{-3}$. If they are present with the coefficients of order one, we can
forget almost everything we said about the predictions, especially in the
Yukawa sector. However, we actually know that the presence of $1/M_{Pl}$
operators is not automatic (at least not with the coefficients of order 1).
Operators of the type (in symbolic notation)
\beq
O_{5}^{p} = \frac{c}{M_{Pl}} 16_{F}^{4}
\eeq
are allowed by SO(10) and they give
\beq
O_{5}^{p} = \frac{c}{M_{Pl}} \left[( Q Q Q L ) + (Q^{c} Q^{c} Q^{c} L^{c}
)\label{30}
\right]
\eeq
These are the well-known $d=5$ proton decay operators, and for $c\simeq O(1)$
they give $\tau_{p}\simeq 10^{23} yr.$  Agreement with experiment requires
\beq
c \leq 10^{-6}
\eeq
Could this be a signal that $1/M_{Pl}$ operators are small in general?
Alternatively, you need to understand why just this one is to be so small. It is
appealing to assume that this may be generic; if so, neglecting $1/M_{Pl}$
contributions in the study of fermion masses and mixings is fully justified.

\subsection{Leptogenesis}

The see-saw mechanism provides a natural framework for baryogenesis through
leptogenesis, obtained by the out-of-equilibrium decay of heavy right-handed
neutrinos\cite{Fukugita:1986hr}. This works nicely for large $M_{R}$, in a sense too
nicely. Already type I see-saw works by itself, but the presence of the type II
term makes things more complicated.
One cannot be a priori sure whether the
decay of right-handed neutrinos or the heavy Higgs triplets is responsible for
the asymmetry, although the hierarchy of Yukawa couplings points towards
$\nu_{R}$ decay. In the type II see-saw, the  most natural scenario is the
$\nu_{R}$ decay, but with the triplets running in the loops
\cite{Hambye:2003ka}.     It appears that the sign of the asymmetry gets correctly predicted
for the type 2 seesaw, an impressive result \cite{babuplanck05}.

\section{Summary and Outlook}

  We have  argued in favor  of SO(10) as the minimal consistent
supersymmetric grand unified theory.  It includes all the interesting features of
Left-Right and Pati-Salam symmetries, it is the ideal setting for a see-saw
mechanism,  and  has the MSSM with automatic R-parity  as the low energy
limit. It can give connections with low energy phenomenology, such as the
one relating $b-\tau$ unification with neutrino mixings, besides being able to
 provide realistic charged fermion spectrum. 

As a gauge symmetry accommodating all fermions of one generation  in a single
representation, there is little doubt on the  convenience of SO(10).  The
question of the Higgs sector is the one unsolved: one can
choose between two different approaches. One can insist
on perturbativity all the way to the Planck scale and choose small
representations, using then   $1/M_{Pl}$ operators to generate the physically acceptable
superpotential; it is then necessary to use textures to simplify the theory.
In this sense, this programme
  appeals to physics beyond grand unification. 
The other approach is to stick to the pure SO(10) theory, at the expense of
using  very large  representations. The couplings then become strong  at 
$\lambda_{F} = 10
M_{GUT}$, but the theory has the advantage of requiring only a
 small number of couplings,  and is a complete theory of matter
and non-gravitational  interactions.  

The important question is rather if these versions of the theory can be tested
in the near future. Work is in progress by several groups on the possibility 
of establishing testable constraints on  neutrino masses and mixings, proton
decay, and the implementation of the leptogenesis scenario.  In the pure
SO(10) approach, with less parameters,  proving the theory wrong might be just  a
question of time.

  \section{Acknowledgements}

We wish to acknowledge many discussions and enjoyable collaboration in the
subjects of these talks with  Charan  Aulakh, Borut Bajc, Thomas Hambye,
 Pavel Fileviez-P\'erez, 
Andrija Ra\v{s}in  and Francesco Vissani. 
  The work of G.S.\ was supported in part by European Commission 
under the RTN contract MRTN-CT-2004-503369; the work of 
A.M.\ by CDCHT-ULA project No.\ C-1244-04-05-B. And we thank the organizers of BW2003, in particular  Goran Djordjevic,
for a stimulating conference and a good time in Vrnja\v{c}ka Banja.
 A. M. wishes
to say hvala, bre. 

{\bf Added note:}

There has been a lot of activity in the subject of supersymmetric SO(10) 
since these talks were presented. In particular, two groups made numerical studies in favor of the
minimal theory discussed here (\cite{Babu:2005ia,Bertolini:2005qb}). The complete study of
proton decay and leptogenesis remain an important challenge.
Meanwhile, the search for the minimal predictive SO(10) theories carried on to the
split-supersymmetric and non-supersymmetric (or even
 supersplit-supersymmetric\cite{Fox:2005yp})
versions, since the gauge coupling unification allows for both. For example, one could get rid of
${\bf 126_H}$ in favor of ${\bf 16_H}$ and the two-loop radiative mechanism for the right-handed
neutrino mass. There are no loops for the superpotential in supersymmetry, and in this case the
ideal scenario becomes split supersymmetry, with light gauginos and higgsinos and heavy fermions
as close to $M_{GUT}$ as allowed by phenomenological constraints \cite{Bajc:2004hr,Bajc:2005aq}. This implies
giving up completely on naturalness, and then one can as well resort to the ordinary,
non-supersymmetric SO(10). For the construction of minimal predictive models in this scenario, see
\cite{bmsv05}.

\end{document}